\documentclass{amsart}
\newtheorem{thm}{Theorem}[section]

\newtheorem{lem}[thm]{Lemma}
\newtheorem{prop}[thm]{Proposition}
\theoremstyle{definition}

\theoremstyle{remark}
\newtheorem{rem}[thm]{Remark}
\numberwithin{equation}{section}

\newcommand{\de}[2]{\frac{\partial #1}{\partial #2}}
\newcommand{\intprod}{\hskip 2pt \vbox{\hbox{\vbox to .18 truecm{\vfill\hbox to .25 truecm
{\hfill\hfill}\vfill}\vrule}\hrule}\hskip 4 pt}

\begin{document}

\title{Quaternionic Hamilton equations}%
\author{Paola Morando}%
\address{Politecnico di Torino, C.so Duca degli Abruzzi 24, 10129 Torino}%
\email{morando@polito.it}%
\author{Massimo Tarallo}%
\address{Universit\`a degli Studi di Milano, Via Saldini 50, 20133 Milano}%
\email{massimo.tarallo@mat.unimi.it}%


\begin{abstract} The classical Hamilton equations are reinterpreted by
means of complex analysis, in a non standard way. This suggests a natural
extension of the Hamilton equations to the quaternionic case, extension
which coincides with the one introduced in \cite{GM1} by a completely
different approach.
\end{abstract}
\maketitle

\section{Introduction}\label{intro}
In the last years, a better understanding of the geometrical framework of the
classical hamiltonian dynamics led to some interesting generalizations.
This is for instance the case of the Nambu dynamics (\cite{Na}, \cite{Ta},
\cite{Mo}), where the evolution of the system in a $n$- dimensional manifold
is related to $(n-1)$ hamiltonian functions, or of the Liouville dynamics,
whose properties are studied in \cite{Ma}.
\\
Very recently, in \cite{GM1} the authors proposed a different extension of the
classical hamiltonian dynamics to spaces which are endowed by three symplectic
structures, having the same commutation rules as the quaternionic imaginary units.
Roughly speaking, this new hyper--hamiltonian dynamics is the sum of three different
hamiltonian dynamics, each one corresponding to a single symplectic structure and to
an arbitrary hamiltonian function. The hyper--hamiltonian dynamics retains most of
the appealing features of the classical hamiltonian one, included a variational formulation
(\cite{GM1}, \cite{GM2}).
\medskip

Here the reader will not be faced with any new application of the hyper--hamilton\-i\-an
dynamics, nor with any discussion about its properties. The aim of this paper is to
point out that the extension given in \cite{GM1}, which is natural from the geometrical
point of view, is also natural from the point of view of analysis. More precisely,
if the classical hamiltonian dynamics is reinterpreted in the light of the complex
analysis, and then extended by replacing the complex analysis with the quaternionic
one, what one gets is again the same hyper--hamiltonian dynamics introduced in \cite{GM1}.
\medskip

\noindent
It should be stressed that the former step does not consist in writing the Hamilton
equations by using complex coordinates, but rather in having the linear notion underlying
complex holomorphy (i.e. complex linearity) coming into play. This is done in section
\ref{complex}.
\\
When the complex numbers are replaced by the quaternions, it is well known that
the complex holomorphy correspondingly changes in the so called \textsl{quaternionic
regularity} (\cite{Fu}, \cite{Su}). This is done in section \ref{completion}, where
the results in \cite{GMT} are used to tune the translation process to our needs.
\\
Finally, in section \ref{hyper} the linear quaternionic regularity replaces
the complex linearity into the equation governing the dynamics. To conclude,
it is shown that the resulting equations coincide with the one already obtained
in \cite{GM1} by a completely different approach.
\\
Section \ref{setting} is devoted to introduce the right context where the
previous program can be worked out. The real needs are metrics and complex
structures, which calls for K\"ahler manifolds instead of just symplectic
one, as it may be expected.
\medskip

\noindent
\textbf{Notations.}\ \ The symbols $\mathbb{C}$ and $\mathbb{H}$
denote the complex numbers and the quaternions; their
standard bases are $1, i$ and $1, i_1, i_2, i_3$ respectively.
Repeated indexes, like in $i_\alpha \partial/\partial x^\alpha$, allude
to a summation on $\alpha=1, 2, 3$; when misunderstandings are possible,
explicit summations or some extra information are added.
Finally, if $X$ is a vector field over a differential manifold $M$, by
$X\;\intprod\; t$ one intends its contraction with the tensor field $t$.
\medskip

\noindent
\textbf{Acknowledgements.}\ \ The authors would like to thank G. Gaeta for many helpful
discussions on the subject.

\section{General setting}\label{setting}
Let consider a Riemannian manifold $M$ with a metric $g$. On $M$,
take a $g$--invariant complex structure $J$, namely a linear map
$TM\to TM$ such that:
$$
J^2=-I\qquad\qquad\textrm{and}\qquad\qquad g(JX,JY)=g(X,Y)\ \ \ \forall X,Y\;.
$$
The manifold $M$ is said to be a K\"ahler manifold when the 2--form
defined by:
\begin{equation}\label{omega}
\omega(X,Y)\ =\ g(JX,Y)
\end{equation}
is closed. Since $\omega$ is nondegenerate and anti--symmetric, it defines
a symplectic structure on $M$. Given a function $h : M\to\mathbb{R}$,
the Hamilton equations may be introduced:
\begin{equation}\label{c-ham}
X\intprod\omega\ =\ dh\;.
\end{equation}
Of course one can define Hamilton equations on \textsl{any} symplectic manifold,
regardless its symplectic structure originates from a metrics or it doesn't.
However, this restriction is essential for all the discussion in the next
section.
\\
In \cite{GM1} the Hamilton equations are extended to cover the case where $M$ is an
hyper--K\"ahler manifold. Instead of a single one, $TM$ has now three
$g$--invariant complex structures such that:
\begin{equation}\label{comm}
J_1^2 = J_2^2 = -I\quad J_1 J_2 = - J_2 J_1\quad J_3 = J_1 J_2\;,
\end{equation}
and all the 2--forms $\omega_\alpha$ associated to $J_\alpha$ via
(\ref{omega}) are closed (see e.g. \cite{Sw}). The hyper--hamiltonian
mechanics introduced in \cite{GM1} then depends on the choice of three
distinct hamiltonian functions $h^\alpha$, its dynamics being defined
by the following equations:
\begin{equation}\label{gm-ham}
X\intprod\Omega\ =\ \frac{1}{(2n-1)!}\;\;
\textstyle{\sum_\alpha} dh^\alpha\ \wedge\ (\omega_\alpha)^{2n-1}\;.
\end{equation}
where $\Omega$ is the volume form on $M$, whose dimension is $4n$, and
$(\omega_\alpha)^{2n-1}$ denotes the wedge product of the differential form
$\omega_\alpha$ by  itself for\ $2n-1\ $ times.
\\
A decomposition formula for the vector field $X$ is then proved in \cite{GM1}:
\begin{equation}\label{dec}
X\ =\ X_1 + X_2 + X_3\qquad\qquad\hbox{where}\qquad\qquad
X_\alpha\ \lrcorner\ \omega_\alpha\ =\ dh^\alpha\quad\forall\alpha\;,
\end{equation}
suggesting the most general hyper--hamiltonian dynamics is the sum of three
independent hamiltonian dynamics.

\section{Hamilton equations and complex analysis}\label{complex}

Consider first classical hamiltonian mechanics. By making use either of the
metrics $g$ and of the complex structure $J$ on $TM$, one can rewrite the
Hamilton equations (\ref{c-ham}) in the form:
$$
g(X,Y)\ =\ dh(JY)\qquad\forall Y\;.
$$
The key point is now to give a closer look at the 1--form lying at the right
hand side, namely:
$$
\theta(Y)\ =\ dh(JY)
$$
from the point of view of complex analysis. First of all, notice that
$M$ is an \textsl{almost complex manifold} in a natural way. Indeed defining
$$
(z Y)_p := t Y_p + x J_p Y_p\qquad\textrm{where}\qquad z=t+ix\in\mathbb{C}
$$
makes $T_p M$ a complex linear space, for every $p\in M$. As a consequence, it
makes sense to talk about complex linear maps on it, and one easily realizes that
$\theta$ is the only \textsl{real valued} 1--form for which the
\textsl{complex valued} 1--form $\ \theta+i\; dh\ $ is in fact a
\textsl{complex linear map}. Such a completion property deserves the ad
hoc notation:
$$
\widehat{i\;dh}\ =\ \theta\;.
$$
The obvious identity $\ i\;dh = d(ih)\ $ then suggests the following
approach. First of all think of the pure imaginary complex function:
\begin{displaymath}
H = i h
\end{displaymath}
as the hamiltonian function describing the system, and then rewrite
the Hamilton equations in the equivalent form:
\begin{equation}\label{cx-ham}
X\intprod g\ =\ \widehat{dH}\;.
\end{equation}
Until now, nothing really new. However, everyone can now guess how
to extend the approach from complex numbers to quaternions.
Assume $M$ is an hyper--K\"ahler manifold, and just think of the
hamiltonian as a function:
$$
H : M\to\mathbb{H}\qquad\qquad H\ =\ h^\alpha i_\alpha
$$
with pure imaginary values (summation on $\alpha$ is implicit). Then think of
$\;\widehat{dH}\;$ as the completion of its differential with respect to the natural
notion of holomorphy in the framework of quaternionic analysis: the so called
regularity (\cite{Su}, \cite{GMT}). This will be done in the next section,
where the following formula:
$$
\widehat{dH}(Y)\ =\ dh^\alpha(J_\alpha Y)
$$
will be proved.
\\
Equation (\ref{cx-ham}) makes sense also in this new framework, and selects
a unique vector field $X$. The important point is that, maybe surprisingly,
$X$ is exactly the same vector field defined by (\ref{gm-ham}). This is the
main result of the paper, and it will be proved in the last section.

\section{Regularity and completion to a regular form}\label{completion}

In 1935 Fueter proposed a definition of \textsl{right regularity} for a function
$\varphi : \mathbb{H}\to\mathbb{H}$, by adapting the Cauchy--Riemann
equations to the increased number of independent imaginary units in $\mathbb{H}$,
namely (\cite{Fu}):
\begin{equation}\label{fue-par}
\de{\varphi}{t} + \de{\varphi}{x^\alpha} i_\alpha  = 0\qquad
q=t+x^\alpha i_\alpha\in\mathbb{H}\;.
\end{equation}
By putting the units $i_\alpha$'s on the left, one obtain the so called
\textsl{left regular} functions, whose theory, clearly, is completely
equivalent to the one of right regular functions.
\\
To a large extent, the quaternionic analysis based on this notion mimics
the richness of complex analysis (see (\cite{Su}), though the non
commutativity of $\mathbb{H}$ introduces some important differences: for
instance, the product or the composition of regular functions need not to be
regular.
\\
Notice that equation (\ref{fue-par}) is in fact a condition on the
differential of $d\varphi$. Indeed, $\varphi$ is right regular if and only
if $d\varphi$ belongs to a special class of 1--forms $T\mathbb{H}\to\mathbb{H}$,
namely the $\theta$'s which satisfy the condition:
\begin{equation}\label{fue-diff}
\theta(1) + \textstyle{\sum_\alpha} \theta(i_\alpha) i_\alpha\ =\ 0\;.
\end{equation}
As pointed out in the introduction, the interest here is essentially in
a more general type of 1--forms, that is:
$$
\theta : TM\to\mathbb{H}\;.
$$
where $M$ is an hyper--K\"ahler manifold. What does it mean regularity in
this framework, is a question which may be answered using the results
in (\cite{GMT}).
\\
In that paper, the notion of regularity is extended to
functions acting between general quaternionic spaces, and the extension is shown
to be strictly dependent on if $\mathbb{H}$ acts on the involved spaces either on
the left or on  the right. For instance, in \cite{GMT} is proved that the above
considered notions of regularity make sense only when $\mathbb{H}$ acts differently
on the two spaces. In particular, the needs for the right regularity are a domain which is a
left space, and a codomain which is a right one.
\\
Now, the way $\mathbb{H}$ acts on the domain $TM$ of $\theta$ is in fact implicit
in our definition of hyper--K\"ahler manifold. Indeed, due to the commutation  rules
(\ref{comm}), the position:
$$
(q Y)_p = t Y_p + x^\alpha (J_\alpha)_p Y_p\qquad\textrm{where}\qquad
q=t+x^\alpha i_\alpha\in\mathbb{H}
$$
makes every $T_p M$ a \textsl{left} quaternionic space. Hence, as pointed out above,
the codomain $\mathbb{H}$ has to be considered as a \textsl{right} space over
itself, and the only applicable notion of regularity is the right one.
\\
Following \cite{GMT}, the correct definition of regularity for $\theta$ is then:
\begin{equation}\label{reg}
\theta(Y) + \textstyle{\sum_\alpha} \theta(i_\alpha Y)i_\alpha = 0
\qquad\forall Y\;.
\end{equation}
That (\ref{reg}) is really an extension of (\ref{fue-diff}) depends on the quaternionic
anti--linearity of its left--hand member (\cite{GMT}). Because of that indeed,
when $\ M=\mathbb{H}\ $ the equality in (\ref{reg}) holds for any $Y\in T\mathbb{H}$ if and
only if it holds for $Y=1$.
\\
Before going further, a remark is in order about the invariance of the
notion of regularity. Looking at (\ref{reg}), this notion seems to depend
on the choice of the imaginary units $i_\alpha$'s. Now, in $\mathbb{H}$
there is an $S^2$ of imaginary units, and any choice of a triple of units,
having the right commutation properties, is an admissible one: a natural
quaternionic object must be independent on that choice. The following lemma
says this is the case for the notion of regularity.
\begin{lem}\label{indip}
If the triple $j_1$, $j_2$ and $j_3$ of quaternions is chosen in such a
way that:
$$
j_1^2 = j_2^2 = -1\qquad\qquad j_3 = j_1 j_2\;,
$$
then, for any quaternionic valued 1--form $\theta$, the following identity
holds:
$$
\textstyle{\sum_\alpha} \theta(i_\alpha Y)i_\alpha\ =\
\textstyle{\sum_\alpha} \theta(j_\alpha Y)j_\alpha \qquad\forall Y\;.
$$
\end{lem}
\proof Define a $3\times 3$ real matrix $C$ by means of:
$$
j_\alpha = C_\alpha^\beta\ i_\beta\qquad\forall\alpha\;.
$$
The commutation rules in the hypothesis say that $C^T C=I$.
The conclusion follows from straightforward computations.\qed
\\\\
In the next proposition the regularity will be exploited to
introduce the notion of completion to a regular form, for 1--forms
$\xi$ which take their values in the imaginary quaternions, namely:
\begin{equation}\label{im-form}
\xi : TM\to\mathbb{H}
\qquad\hbox{with}\qquad\xi+\overline{\xi}=0\;.
\end{equation}
\begin{prop}\label{compl}
If $\xi$ satisfies condition (\ref{im-form}), then there is a unique
real valued 1--form $\widehat{\xi}\in T^* M$ such that $\ \widehat{\xi} + \xi\ $ is a
regular 1--form. Moreover, the completion $\ \xi\mapsto\widehat{\xi}\ $ is a real linear
operation and, if $\ \xi\;=\;\xi^\alpha i_\alpha\ $ with $\ \xi^\alpha\in T^*M$, then:
$$
\widehat{\xi}(Y) = \xi^\alpha(i_\alpha Y)\qquad\forall Y\;.
$$
\end{prop}
\noindent
The proof is a straightforward consequence of the following lemma,
where regularity of the general 1--form $\theta$ is rewritten using
its components.
\begin{lem}\label{reg-comp}
Assume $\theta = \theta^0 + \theta^\alpha i_\alpha$, where $\theta^0$ and
each $\theta^\alpha$ lie in $T^*M$ (namely are real valued 1--forms).
Then $\theta$ is regular if and only if:
$$
\theta^0 (Y)\ =\ \theta^\alpha (i_\alpha Y)\qquad\forall Y\;.
$$
\end{lem}
\proof Just write down the sum in (\ref{reg}), separating the
contributions along $1$ and each $i_\alpha$. The four conditions
obtained in this way are all equivalent: the statement expresses the
vanishing along $1$. \qed

\begin{rem}\label{indip2}
Like the notion of regularity, the notion of completion is also
independent on the choice of the imaginary units $i_\alpha$.
\end{rem}

\section{Quaternionic Hamilton equations}\label{hyper}
Coming back to dynamics, take an hamiltonian function:
$$
H : M\to\mathbb{H}\qquad\qquad\mathrm{with}\qquad\qquad \overline{H} + H = 0\;.
$$
on the hyper--K\"ahler manifold $M$. Proposition \ref{compl} applies
to the 1--form $dH$, and its completion may be used to introduce the Hamilton like
equation:
\begin{equation}\label{q-ham}
X\intprod g\ =\ \widehat{dH}\;.
\end{equation}
\begin{rem}
In some special situations, the quaternionic Hamilton equations coincide
with the classical ones. This is the case when the hamiltonian function
is directed along a single pure imaginary quaternion, namely:
$$
H(p) = h(p)\ u\qquad\qquad h\ \hbox{real valued,}\quad u^2=-1
$$
Indeed, by means of Proposition \ref{compl} and the remark thereafter:
$$
\widehat{dH}(Y)\ =\ dh(uY)\qquad\forall Y,
$$
so that equation (\ref{q-ham}) becomes:
$$
X\intprod\omega_u\ =\ dh\;.
$$
Here, of course, $\omega_u$ is the symplectic form associated
to the complex structure $J_u Y = u Y$.
\end{rem}
\noindent
To complete the program announced in the introduction, it remains to show
that both the equations (\ref{gm-ham}) and (\ref{q-ham}) define the same
vector field $X$. This is a consequence of the following result and formula
\ref{dec}.
\begin{prop}\label{sym}
Assume $\ H = h^\alpha i_\alpha$. Then the vector field $X$ satisfies
equation (\ref{q-ham}) if and only if it decomposes as:
$$
X = X_1 + X_2 + X_3
$$
where:
$$
X_\alpha\intprod\omega_\alpha\ =\ dh^\alpha\qquad\forall\alpha\;.
$$
\end{prop}
\proof
Using Proposition \ref{compl}, condition (\ref{q-ham}) rewrites as:
$$
g(X,Y) = dh^\alpha ( i_\alpha Y)\qquad\forall Y\;.
$$
The decomposition suggested in the statement clearly gives rise to a
a vector field $X$ which satisfies the previous condition.
On the other hand, there is only one vector field which has this property.
\qed
\\\\
Summing up, two possible extensions of classical hamiltonian mechanics
have been considered. One is the hyper--hamiltonian mechanics introduced in
\cite{GM1}, whose dynamics is defined by (\ref{gm-ham}), and which is natural
from the point of view of symplectic geometry.
The other one is the quaternionic hamiltonian mechanics introduced
in this note, whose dynamics is defined by (\ref{q-ham}), and which is
natural from the point of view of complex and quaternionic analysis.
Though they arise from apparently unrelated approaches, the two extension
coincide.
\\
Moreover, in the quaternionic approach the exact meaning of the sum in
(\ref{dec}) also becomes clear. The intrinsic object is the hamiltonian
function $H$, whereas the $h^\alpha$ are just its components with respect
to a given basis in the space of imaginary quaternions. Changing the base,
the components of $H$ change, and so do the vector fields $X_\alpha$:
however, its sum is intrinsic inasmuch as it does not depend on any choice.


\end{document}